# The Obvious Solution to Semantic Mapping – Ask an Expert

Kieran Greer, Distributed Computing Systems, Belfast, UK.
http://distributedcomputingsystems.co.uk
Version 1.1

***Abstract:*** The semantic mapping problem is probably the main obstacle to computer-to-computer communication. If computer A knows that its concept X is the same as computer B's concept Y, then the two machines can communicate. They will in effect be talking the same language. This paper describes a relatively straightforward way of enhancing the semantic descriptions of Web Service interfaces by using online sources of keyword definitions. Method interface descriptions can be enhanced using these standard dictionary definitions. Because the generated metadata is now standardised, this means that any other computer that has access to the same source, or understands standard language concepts, can now understand the description. This helps to remove a lot of the heterogeneity that would otherwise build up though humans creating their own descriptions independently of each other. The description comes in the form of an XML script that can be retrieved and read through the Web Service interface itself. An additional use for these scripts would be for adding descriptions in different languages, which would mean that human users that speak a different language would also understand what the service was about.

## 1   Introduction

Semantic mapping is probably the main impediment to the Semantic Web [2][7] at the moment. Semantic mapping means that if computer A is looking for a concept, such as a 'car', and computer B is providing 'automobiles', through mapping these concepts together, or determining that they are the same, both computers can communicate about the request. With their lack of intelligence, computers need very simple and foolproof ways of determining if two words have the same meaning. Typically, some form of statistical process is used that matches words based on how similar their letter structure is [3]. There are also ontologies [7] that can be referenced to determine, for example, that a car is an automobile. But for these to work the computers need to be using the same set of concepts to define everything, and as they would typically be built up independently of each other, there is no guarantee that this would be the case. If the mapping problem can be overcome however, then we are at least half way to the Semantic Web. The point being that either computer should understand what its own concepts represent and so if it can safely map these to the other computer's concepts, it will know what they represent also. Computers will then talk the same language and be able to communicate with each other. This will allow for many different forms of autonomic communication and processing, from asking a search agent to search the internet for your best holiday bargain, to two computer systems having an intelligent conversation and reasoning about the answers that each receives. An application is described in this paper that allows a user to generate more detailed and standardised descriptions of the web service interface, increasing the chances of such conversations taking place.

The first hurdle to overcome is to allow one computer program to invoke a service or action on another computer program. Over the Internet, Web Services [5] offer a standard communication protocol that can be used to invoke service operations. Web Services are often defined by standardised XML-based scripts, such as WSDL. This essentially provides descriptions of where the Web service is and the methods that can be invoked on it. With the more abstract RESTFul Web Services [9] however, there might not be a full WSDL description, when the calling program would then need additional help to know what the service interface represents. One problem with this is the fact that often programmers do not use conventional words to define their parameters and the syntax is nearly always very heterogeneous. There are however published coding standards that help to provide a uniform structure to the method or parameter definitions. For example, a parameter called 'getCarType' can be reliably parsed into the three words 'get', 'car' and 'type', which can be determined by each new word starting with a capital letter. While some knowledge and skill is required to determine which of these keywords and definitions to then use, the application described in this paper largely automates the process, making it much easier for the writer of the service to do this. It then becomes more of a multiple choice question, rather than requiring the user to create a definition from scratch. The other advantage of the mechanism that will be described is the fact that the descriptions are based on standard definitions retrieved from online dictionaries. These can even be checked if the web address is known. This is therefore a ready-made standard for defining your web service interface that can be utilised relatively easily.

## 2 Mapping Metadata and Semantics

Computer systems still need help to overcome their limitations. The main problem at the moment is what it has always been - they are still not intelligent enough. They do not understand natural language and are not able to understand each other. There is very little generality and they cannot reason outside of very restricted domains of knowledge. Because of this they require some help if they are going to be able to understand each other. Cloud computing [1] is the new architecture for providing services over the Internet. It is similar to Grid or SOA with the principle feature of a distributed Internet-driven architecture for providing these services, and all of these architectures would benefit from adding autonomous or intelligent behaviour to the computer programs that manage and use them. To allow them to perform more intelligent tasks, the users or writers of the systems need to provide knowledge to the system in a structured and standardised way. Ontologies [6][7] provide a suitable structure for representing such knowledge. Unfortunately however, ontologies are generally built up in isolation of each other and so when the concepts in each are compared there is some degree of heterogeneity that the computer can still not robustly deal with. The hierarchical structure in the ontology can be used to tell if different concepts are the same and statistical processes can compare words directly to tell how similar they are. Experience-based techniques can now also be used, for example, the context in which the concepts are used. But without any real intelligence, the key at the moment will be to provide sufficiently detailed metadata describing the concepts. This will broaden the scope of the semantic and syntactic matching, thus increasing the chance that a suitable match can be found. For example, consider the following two sentences:

- There is a car in the garage.
- There are cars in the car park.

The computer has to determine if 'car' and 'cars' are the same concept. Statistical matching indicates only 1 difference in the two words and so with some degree of confidence the computer can guess that the two concepts are the same. What if the computer needs to find a suitable mechanic to service a car for MOT? Maybe two sites advertise their services as:

1. I can service your vehicle for MOT.
2. I provide a service of finding classical cars for purchase that have passed their MOT.

In this case both descriptions have the word 'service', but the meaning is very different. So instead of matching over just the words 'service' and 'car', the system can try to match over the full description. The first sentence is missing the car keyword, but has the vehicle keyword. The web interfaces have provided additional metadata to describe the keywords in their descriptions, where vehicle can be described as:

- Vehicle - a car, lorry, bus, etc., for transporting people or goods on land.

So matching to this then gives a match to car as well. The second service looks quite promising as well, as keywords such as classical might not be considered to be relevant. But maybe the service keyword here is described as:

- Service – help or advice.

The service description we are looking for is something like:

- Service – a routine inspection and maintenance of a vehicle.

As these two 'service' descriptions differ a lot, it is unlikely that the second site does in fact provide the required service. Putting this together, the agent would opt for the first site over the second one for the following reasons:

- Vehicle can be directly mapped to car.
- Service has the correct description.

It is also clear however that some time and effort is needed to further define and describe key concepts to allow the statistical or ontological mapping techniques to work. Online dictionaries can be used to provide these standard descriptions for any concept and by using standard ontology descriptions, at least the heterogeneity in the concept description will be removed.

## 3  Metadata and Semantic Descriptions

The example of section 2 is the sort of information that you need to provide when describing the methods that you expose on your Web Service interface. The Code Analyzer application that is described in this section provides the capabilities for adding this sort of descriptive metadata to your service. The graphic of Figure 1 shows what the metadata panel looks like, with a code file parsed into keywords displayed, together with some additional metadata descriptions.

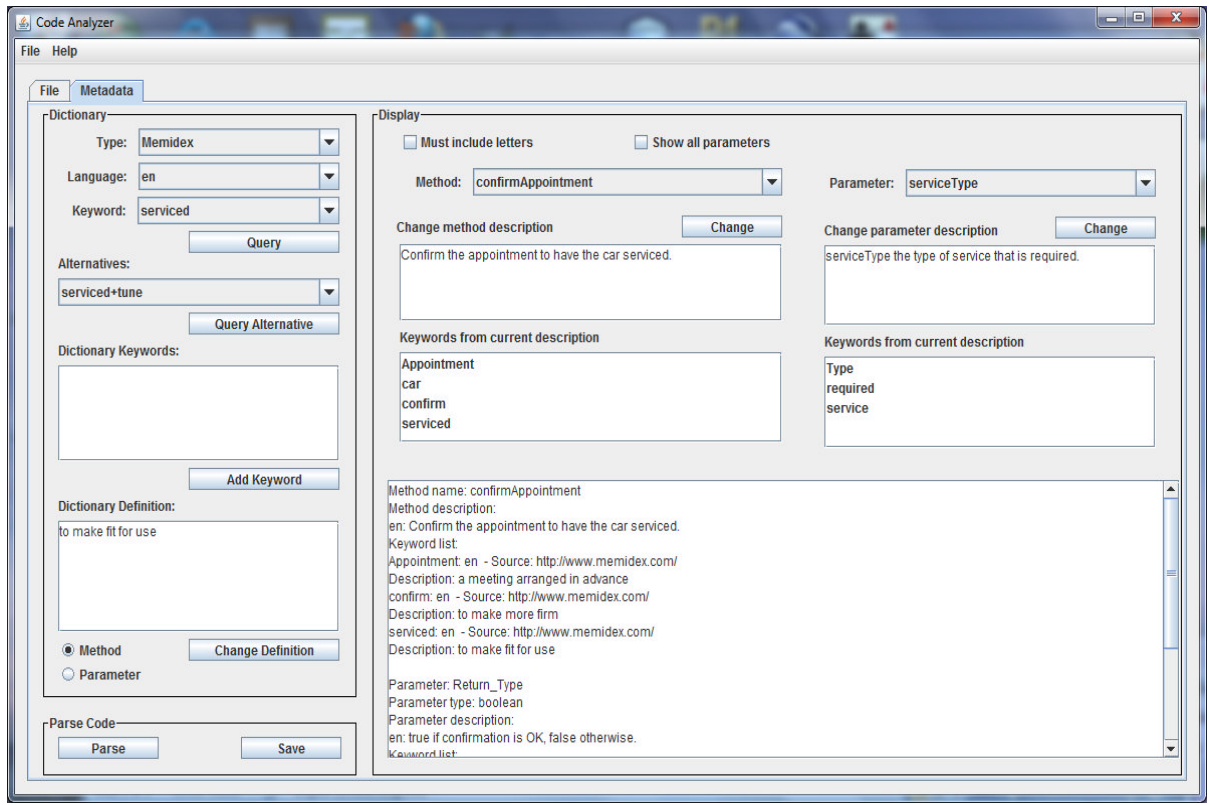

Figure 1: Semantic descriptions panel showing parsed code file and added descriptions.

The file type is defined by its extension type. A demo version of the application can be downloaded from the sourceforge.net web site [4]. The demo version currently supports the following types of code file:

- A WSDL file is defined with a .wsdl or .xml extension.
- A Java file is defined with a .java extension.

The method declarations in the code file are parsed and tokenised to generate a number of keywords that can be queried on an online dictionary for standard meanings and definitions. The bottom right text area shows the current metadata description of the code in a string-based format. When you save this it is converted into an XML file, but the string-based representation is easier to read for display purposes. The left-hand side of the panel shows how an online dictionary can be selected and a chosen keyword queried on it. There is also the possibility of selecting the language to query, depending on what online source is chosen. The demo version comes with three different online dictionaries. These are:

- *FreeDicts*: url - http://www.dicts.info/
- *Memidex*: url - http://www.memidex.com/
- *SynonymsDict*: url - http://www.synonym.com/

Each dictionary might return different information and so when a definition is used, its source URL is also stored with it. There are selection buttons to indicate if you are updating the method or parameter's description. In this case the method's metadata is being looked at and the selected keyword is 'serviced'. Note that the keyword list is not case sensitive.

You can add keyword lists with definitions to each parameter or method, to further enrich its description. Each new description that is added is additional to the existing metadata. The new metadata is then displayed as a keyword with source and definition as follows:

- The keyword added is the one in the 'Keyword' combo box, which is the one that was originally queried.
- The source is the URL of the online dictionary.
- The language is the language code in the 'Language' combo box. This is only allowed to be 'en' (English) in the demo version.
- The definition is the definition returned by the query, in the 'Dictionary Definition' text area.

Whenever you have added sufficient metadata descriptions to all of your methods and parameters you can save the description as an XML script. This script or metadata description can then be read by another program. A second jar file is provided with the application that you can add to your web service interface and it will read the generated script and return the relevant part relating to any method or parameter name. The retrieved information will be the stored definitions, and lists of keywords with related meanings. This provides a richer and more standardised information source with which to make semantic comparisons.

# 4    Pros and Cons

This is actually a very obvious way to try and solve the semantic mapping problem, but it does not appear to be widespread even though it looks relatively straightforward and useful. So what would the reasons be for not adopting this mechanism?

## 4.1    Cons

Why has this sort of mechanism not become commonplace? For one thing programmers now like to be able to automatically create their Web service through an integrated IDE. For this they simply define the method requirements and the programming environment does the rest. It would be difficult to incorporate this sort of functionality into an IDE as standard. It also requires a certain level of additional thinking and some additional effort to actually query and select from a database the appropriate terms and definitions that will provide the extra quality to the method definition. As with other attempts to standardise these interfaces, it also requires a widespread adoption for a useful number of systems to be able to query each other using the mechanism. Also, just because there is more descriptive information does not automatically mean that there will be a better match. Some of the additional content could be misleading, for example. The additional content needs to be accurate and standard, and so for the moment, a research environment might be preferred for testing this mechanism.

## 4.2    Pros

While this application might not be part of an IDE, it is relatively easy to use. The dictionaries are all online and so with just one or two clicks you can retrieve the information you require. While you still have to think about keywords and intended meanings, etc., the system provides you with possibilities to choose from. This means that it is more of a multiple choice question and there is no requirement to create a new word or description yourself. Once you have taken the time to add the additional metadata to your service, it is a standard definition and so this should really help if you want to make your service part of some sort of autonomous system (grid or cloud). The script itself is very easy to generate and use, while the same procedure could be applied for adding it to any service interface.

## 5   Conclusions

If the goal is to add autonomous or agent-based behaviours to your system, then additional descriptive metadata makes it easier for other programs to determine what your program is about. This could also be an additional help for foreign-speaking users, who could retrieve the description of your service in their own language. Adding more metadata potentially also makes the problem worse however, as there is now more to map to; but if the definitions are standard and accurate, then the calling program will be able to map more accurately. As with other attempts to standardise these interfaces, it also requires a widespread adoption for a useful number of systems to be able to query each other using the mechanism. So for the moment a research environment might be preferred. A demo version of the application is currently available and a professional version should be released sometime in the future.